\documentclass[12pt]{iopart}
\usepackage{epsf}
\usepackage{graphicx}
\newcommand{\st}{\sigma_{los}}
\newcommand{\si}{\sigma_{m}}

\def\gtsim {\lower .1ex\hbox{\rlap{\raise .6ex\hbox{\hskip .3ex
        {\ifmmode{\scriptscriptstyle >}\else
                {$\scriptscriptstyle >$}\fi}}}
        \kern -.4ex{\ifmmode{\scriptscriptstyle \sim}\else
                {$\scriptscriptstyle\sim$}\fi}}}

\def\ltsim {\lower .1ex\hbox{\rlap{\raise .8ex\hbox{\hskip .3ex
        {\ifmmode{\scriptscriptstyle <}\else
                {$\scriptscriptstyle <$}\fi}}}
        \kern -.4ex{\ifmmode{\scriptscriptstyle \sim}\else
                {$\scriptscriptstyle\sim$}\fi}}}
                
\begin{document} 

\title[Kinematics of Milky Way Satellites]
{Kinematics of Milky Way Satellites: Mass Estimates, Rotation Limits, and Proper Motions}
\author{Louis E. Strigari} 
\address{Kavli Institute for Particle Astrophysics and Cosmology, 
Stanford University, Stanford, CA 94305, USA}
\ead{strigari@stanford.edu}

\begin{abstract} 
In the past several years high resolution kinematic data sets from Milky Way satellite
galaxies have confirmed earlier indications that these systems are dark matter dominated objects. 
Further understanding of what these galaxies reveal about cosmology 
and the small scale structure of dark matter relies in large part on a more detailed 
interpretation of their internal kinematics. This article discusses a likelihood formalism 
that extracts important quantities from the kinematic data, 
including  the amplitude of rotation, proper motion, 
and the mass distribution. In the simplest model the projected error on the rotational amplitude 
is shown to be $\sim 0.5 $ km s$^{-1}$ with 
$\sim 10^3$ stars from either classical or ultra-faint satellites.  
The galaxy Sculptor is analyzed for the presence of a rotational signal; 
no significant detection of rotation is found, and given this result 
limits are derived on the Sculptor proper motion. 
A criteria for model selection is discussed  
that determines the parameters required to describe the dark matter halo density profiles and
the stellar velocity anisotropy. Applied to four data sets with a wide range of
velocities, the likelihood is found to be more sensitive to variations in the
slope of the dark matter density profile than variations in the velocity anisotropy.  
Models with variable radial velocity anisotropy are shown to be preferred relative to those 
in which this quantity is constant at all radii in the galaxy.  
\end{abstract} 

\maketitle 

\section{Introduction} 
Since their initial discovery~\cite{Shapley}, dwarf spheroidals (dSphs)
have offered a unique insight into the formation of galaxies and structure on the 
smallest scales. Initially characterized as unusual and ghostly stellar systems, photometric 
studies tended to find that these systems contained old stellar populations with no recent signature of star formation 
activity~\cite{Mateo:1998wg}. Though photometrically well-studied since their discovery
over seventy years ago, as late as nearly 30 years ago  
minimal was known on the internal kinematic properties of their stellar populations
or on the kinematic properties of these objects in the Milky Way (MW) halo. 

Aaronson \cite{Aaronson1983} provided the first measurement of the line-of-sight velocities 
of stars in Milky Way dSphs. From the spectra of merely 
three carbon stars, Aaronson suggested a mass-to-light
ratio for the Draco dSph nearly an order of magnitude greater than that of Galactic globular
clusters. Follow-up studies of several dSphs, including Sextans, Fornax, Ursa Minor, 
Sculptor, increased the velocity samples by an order of magnitude, and in the
process established these systems to be dark matter dominated
~\cite{Suntzeff1993,Mateo1991,Mateo1993}. 
It was further suggested that all of these systems share a similar dark matter halo 
mass of $\sim [1-5] \times 10^7$ M$_\odot$~\cite{Mateo1993}. Even at the time of 
these early measurements, it was understood that the mass distributions of 
these systems provide strong constraints on the properties of the particle nature of 
dark matter, including its mass and primordial phase-space density
~\cite{Lake1990,Gerhard1992,Faber1983}. 

With the advent of high resolution, multi-object spectroscopy, the velocity samples
from the brightest dSphs initially studied in Refs.~\cite{Suntzeff1993,Mateo1991,Mateo1993} 
have now increased by up to three orders of magnitude~\cite{Gilmore2007,Walker2007,Walker:2008ji}. 
These new data sets have revealed that the velocity dispersions of the systems are all
$\sim 10$ km s$^{-1}$, and in all cases the dispersions remain constant even out
to the projected radius of the outermost velocity measurements 
~\cite{Walker2007}. 
Though the data sets have increased by more than ten-fold, the more modern analysis of 
these systems still confirms the global conclusion established from the initial observations that
dSphs are strongly dark matter dominated
~\cite{Gilmore2007,Walker2007,Strigari:2008ib,Lokas:2009cp}.  

Not only has the past several years seen an increase in the kinematic data sets for
the brightest dSphs, the number of
known Milky Way satellites has more than doubled 
due to the Sloan Digital Sky Survey (SDSS). As of the writing of this article, 
the SDSS has discovered 14 new Galactic satellites
~\cite{Willman:2004kk,Willman:2005cd,Belokurov:2006ph}. The new SDSS systems
have lower luminosities and surface brightnesses than the 11 classical Milky Way satellites
that were known prior to SDSS. 
The half-light radius for several of these new objects is less than $\sim 100$ pc; this radius
is smaller than the typical half-light radius of the classical satellites but still somewhat larger than
the typical globular cluster half-light radius of $\sim 1-10$ pc. 

Several kinematics studies on the ultra-faint population of SDSS satellites have been 
undertaken in the past several years~\cite{Munoz:2006vg,Martin:2007ic,Simon:2007dq,Geha:2008zr}. 
Using spectra from eight of the SDSS satellites,  Simon and Geha~\cite{Simon:2007dq} 
concluded that these objects are strongly dark matter dominated. Several of the ultra-faint
satellites have velocity dispersions as low as $\sim 5 $ km s$^{-1}$, making them the
most-promising systems to study the phase space limits of the dark matter. It has 
additionally been observed that the ultra-faint satellites are the most metal-poor systems
known, and that they form a continuation of the luminosity metallicity trend set by the brightest dSphs
~\cite{Kirby2008,Geha:2008zr}. 

With the above data sets now available, 
it is becoming increasingly necessary to develop better theoretical tools to 
interpret them. An important aspect of the theoretical modeling will necessarily 
require an interpretation of the kinematic data sets for the population of MW satellites; a  
detailed understanding of these kinematic data sets 
will be important not only for determining the mass distributions of each individual 
system, but for a global comparison to theories of Cold Dark Matter (CDM) 
~\cite{Strigari:2007ma,Strigari:2008ib}. Understanding the mass distributions
will also be important for interpretation of limits on particle dark matter masses and annihilation
cross sections in high-energy gamma-ray experiments~\cite{Strigari:2006rd,Essig:2009jx,Martinez:2009jh}. 
Further, understanding the kinematics of these systems may eventually reveal whether they
have dark matter cusps or cores, which would in itself provide a stringent test of 
the CDM paradigm~\cite{Hogan:2000bv}. 

The primary aim of this article is to discuss a maximum likelihood formalism 
that is used for extracting important physical quantities from dSph kinematic data sets. 
Section~\ref{sec:data} begins by reviewing the properties of the kinematic data sets
and defining the likelihood. Section~\ref{sec:tracers} then uses the likelihood to extract
rotational and proper motion signals. 
Section~\ref{sec:mass} discusses mass modeling and
a new calculation for model selection. 
Section~\ref{sec:conclusion} presents the conclusions. 

\section{Likelihood Function and Error Modeling}
\label{sec:data}
Information on the kinematic properties of dSphs is extracted from the line-of-sight 
velocities of their individual stars. This section introduces the likelihood used in
the data analysis and projections for the errors attainable on several parameters 
using the likelihood. 

\subsection{Likelihood Function} 
The probability for a velocity data set, $\vec v$, is assumed to be of the form
\begin{equation}
p(\vec v | u, \sigma_{los}) = \prod_{\imath=1}^n \frac{1}{\sqrt{2\pi(\sigma_{m,i}^2+\sigma_{los}^2)}}
\exp \left[ -\frac{ (v_\imath - u)^2} {2(\sigma_{m,i}^2+\sigma_{los}^2)} \right ]. 
\label{eq:gaussian}
\end{equation} 
In Eq.~\ref{eq:gaussian} the dispersion of the distribution is given by the sum of the 
measurement uncertainty on a star, $\sigma_m$, and the intrinsic dispersion of the 
system at the projected radius of the star. 
The latter quantity is symbolized by $\sigma_{los}$ and is determined by the  
model; Section~\ref{sec:mass} below provides more details on this quantity and specifically how
it relates to the mass of the systems. 
The systemic line-of-sight velocity in the direction of the $i^{th}$ star is given by $u$.
Written in the above form, Eq.~\ref{eq:gaussian} may be read as the probability for the data set, 
given the parameters $u$ and $\sigma_{los}$. Appealing to Bayes' Theorem and 
defining the likelihood function as
\begin{equation} 
{\cal L}(u ,\sigma_{los}) = p(u, \sigma_{los} | \vec v)
\label{eq:likelihood}, 
\end{equation} 
the parameters $u$ and $\sigma_{los}$ may be determined 
directly from the data by the maximization of Eq.~\ref{eq:likelihood}. 
Equation~\ref{eq:likelihood} assumes uniform priors on the model parameters. 

The form of Eq.~\ref{eq:gaussian} results from the convolution of Gaussian 
distribution which represents the measurement error on the velocity of a
given star with a separate sampling distribution that is assumed to be Gaussian. 
It is the sampling distribution of velocities that is connected to physical quantities such as 
the velocity anisotropy of the stars, and the potential of the stellar and dark matter
components. For a given model of the galaxy, the true line-of-sight velocity distribution function 
may indeed be non-aussian; certain limiting cases of the velocity distribution for analytic potentials 
have been considered in Ref.~\cite{Gerhard1991}. This paper shows that when attempting to reconstruct the 
line-of-sight velocity distribution for a given model, degeneracies exist between the 
stellar velocity anisotropy and the stellar and dark matter potentials. Though more 
information may be gained on model parameters if the true velocity distribution were known,
and thus utilized in the parameter estimation, 
the Gaussian approximation provides the most conservative sampling distribution
in reconstructing model parameters in variance estimation problems (for a specific discussion of this point, see
the discussion in Chapter 8 of Ref.~\cite{Gregory2005}). Further, the mass estimations
presented here using the likelihood in Eq.~\ref{eq:likelihood} agree with mass estimates that use a 
Gaussian likelihood in the binned velocity dispersion
~\cite{Strigari:2007at,KS:unpub}; in this latter case the velocity dispersion does not necessarily 
correspond to the variance of a Gaussian line-of-sight velocity distribution, making it self-consistent to determine
parameters such as the velocity anisotropy. 

The distribution function in Eq.~\ref{eq:gaussian} provides the simplest 
description of a data set. Including higher-order effects naturally introduces a larger 
set of model parameters. The first modification to Eq.~\ref{eq:gaussian}
from higher order corrections comes from noting that 
the mean velocity, $u$, varies as a function of the position of the star in the galaxy. 
This variation in the mean velocity results from the fact that, for lines-of-sight 
with larger angles from the line-of-sight directly to the center of the galaxy, 
the proper motion of the object contributes an increasingly larger component
to the line-of-sight velocity. 
To describe how the line-of-sight velocity varies as a function of position, 
consider a cartesian coordinate system in which the $z$-axis
points in the direction of the observer from the center of the galaxy,
the $x$-axis points in  the direction of decreasing right ascension,
and the $y$-axis  points in the direction of increasing declination.  
The angle $\phi$ is measured counter-clockwise from the positive
$x$-axis, and $\rho$ is the angular separation from the center of the
galaxy.  The mean line-of-sight velocity is then  
\begin{equation} 
u = v_x \sin \rho \cos \phi + v_y \sin \rho \sin \phi
- v_z \cos \rho. 
\label{eq:vlos}
\end{equation}
In the small angle approximation, 
$\sin \rho \simeq R/D$, where $R = \sqrt{x^2 + y^2}$, 
and $D$ is the distance from the observer to the center of the dSph. 
Then $\sin \phi = y/R$, so that 
Eq.~\ref{eq:vlos} can be  written as $u = v_x x/D + v_y y/D -
v_z$.
In the limit that the vector pointing from the observer to the center of the
galaxy is exactly parallel to the lines-of-sight to each star, $u \simeq -v_z$. 

Equation~\ref{eq:vlos} show that the line-of-sight velocity of a system
increases roughly linearly with the increase of the projected distance from the center of
the dSph. This effect is purely geometric and
may be used to recover the proper motion of a dSph with
similar accuracy to the proper motions attained in ground and space-based measurements
~\cite{Kaplinghat:2008sm,Walker:2008ji}; an application to a specific data set of Sculptor is
given below. The extraction of dSph proper motions in 
this manner is analogous to the determination of the proper motions for the Large
Magellanic Clould~\cite{vanderMarel:2002kq} and for M31
~\cite{vanderMarel:2007yw} from their stellar and satellite distributions, respectively. 

There may also be rotational motion, in addition to the dominant 
contribution from random motions, present in the galaxy. Though 
rotation is intrinsic to the dynamics of the system and is 
not purely geometric as that described by Eq.~\ref{eq:vlos}, a simple 
parameterization is possible if the rotation amplitude is described
by a term $A \sin (\phi-\phi_0)$, where $\phi_0$ defines the projected axis
of rotation. Adding all of the terms together gives the following expression for
the line-of-sight velocity of a star,  
\begin{equation} 
u = v_x \sin \rho \cos \phi + v_y \sin \rho \sin \phi
- v_z \cos \rho + A\sin(\phi-\phi_0). 
\label{eq:vlos_A}
\end{equation} 

With the addition of each of the terms in Eq.~\ref{eq:vlos_A}, our likelihood
function now reads
\begin{equation} 
{\cal L}(v_x,v_y,v_z,A,\phi_0, \sigma_{los}) = p(v_x,v_y,v_z,A,\phi_0, \sigma_{los} | \vec v)
\label{eq:likelihood_A}, 
\end{equation} 
and the vector set of 6 parameters ($v_x,v_y,v_z,A,\phi_0, \sigma_{los}$) may be directly
determined from the data. In the sections below these parameters are determined from 
an example data set; before jumping into this data analysis the following sub-section 
provides a discussion of the 
theoretical predictions for the errors attainable on these quantities. 

\subsection{Error Modeling} 
From the likelihood function defined in Eqs.~\ref{eq:gaussian} and~\ref{eq:likelihood_A}, 
the Fisher matrix formalism may be used to derive projected errors on the model 
parameters. For $m$ model parameters that are varied, the Fisher matrix is defined as 
an $m$ by $m$ matrix so that the entry for the $a^{th}$ and $b^{th}$ parameters is
given by
\begin{equation} 
F_{ab} 
= -\left \langle \frac{\partial^2 \ln {\cal L}}{\partial \theta_a \partial \theta_b} \right \rangle.  
\label{eq:fisher} 
\end{equation} 
Here $\vec \theta$ is a vector defining the set of parameters. In the simple 
case studied in this section, the parameters are given by 
$\vec \theta = \{v_x,v_y,v_z,A,\phi_0, \sigma_{los}\}$. 
According to the Rao-Cramer inequality, the minimum 
possible variance attainable on a parameter using maximum
likelihood statistics is given by the inverse of the Fisher 
information matrix, $\sqrt{{\bf F}_{aa}^{-1}}$. 
The average in Eq.~\ref{eq:fisher} is taken over the data, 
and the derivatives are 
evaluated at the true model of parameter space.
The inverse of the Fisher matrix thus 
provides an approximation for the true covariance of the parameters, 
and using ${\bf F}^{-1}$ provides a good approximation to the errors 
on parameters that are well-constrained by the data.  

The Fisher matrix is constructed by differentiating the log 
of the likelihood function in Eq.~\ref{eq:likelihood_A}. 
It will be understood that the total dispersion 
$\sigma_\imath^2 = \st^2 + \si^2$ is 
evaluated at the projected radius of the $\imath^{th}$ star. 
Averaging over the likelihood function, and using the above
definition of $u$,
the final expression for the Fisher matrix is  
\begin{equation} 
F_{ab} 
= \sum_{\imath=1}^{N} \left(
\frac{1}{\sigma_\imath^2}\frac{\partial u_\imath}{\partial \theta_a}
\frac{\partial u_\imath}{\partial \theta_b}
+\frac{1}{2}
\frac{1}{\sigma_\imath^4}\frac{\partial \st^2}{\partial \theta_a}
\frac{\partial \st^2}{\partial \theta_b}\right).  
\label{eq:fisher_averaged} 
\end{equation} 
The sum is over the $N$ number of observed stars in the galaxy. The 
analysis in this section considers the simplified case that $\st^2$ does 
not in itself depend on any model parameters. A more detailed model
would consider this quantity as a function of the parameters that 
describe the mass modeling of the system; this is discussed in more
detail in Section~\ref{sec:mass} below.  

In the second term in Eq.~\ref{eq:fisher_averaged}, the derivatives 
are with respect to the theory dispersion alone, whereas both of the contributions to the 
variance sum in the denominator. For the
well-studied satellites, with intrinsic velocity dispersions of 10 km s$^{-1}$, 
the dispersion from the distribution function dominates the 
dispersion from the measurement uncertainty, while for many of
the newly-discovered satellites, both contributions to 
the dispersions are similar. 
Equation~\ref{eq:fisher_averaged} shows that, to determine the error 
on any of the $\vec{\theta}$ parameters, one must determine
1) the distribution of stars within the 
dSph that have measured velocities, and 2) the error on the 
velocity of each star. This implies that the projected errors are independent 
of the mean velocity of the stars. Additionally, under 
the approximation that $\sin \rho \ll 1$ and no rotation, the first term in 
Eq.~\ref{eq:fisher_averaged} vanishes, and the errors are
independent of the parameters describing the mean 
motion of the system. 

\begin{figure}[htbp]
\begin{center}
\includegraphics[height=8cm]{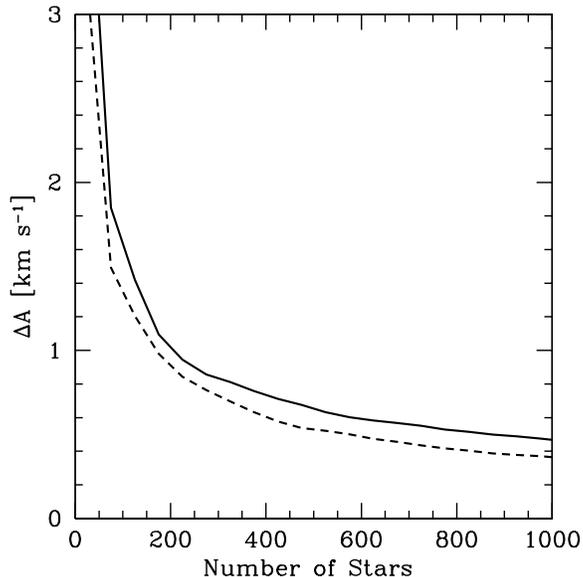}
\caption{The projected one-sigma error on the amplitude of the rotation parameter, $A$,
as defined in Eq.~\ref{eq:vlos_A}.  The upper solid curve assumes structural
parameters similar to that of Segue 1, while the lower dashed curve assumes
structural parameters similar to that of Draco. Each curve assumes that the
measurement uncertainty on each star is 2 km s$^{-1}$. 
\label{fig:rotation}
}
\end{center}
\end{figure}

The projected errors obtained using Eq.~\ref{eq:fisher_averaged} provide 
an excellent estimate of the measured errors on both $v_x$ and $v_y$~\cite{Kaplinghat:2008sm,Walker:2008ji}. 
Though there has been no conclusive detection of a parameter similar to $A$ in 
published kinematic data samples, 
it is interesting to determine the expected error on this quantity given expected future 
data samples. Figure~\ref{fig:rotation} shows
example error projections for $A$, for two different model galaxies. The upper solid curve 
assumes structural parameters similar to that of Segue 1, with a Plummer radius 
of $0.03$ kpc and a stellar limiting radius of $0.1$ kpc~\cite{Martin:2008wj}.
The lower dashed curve 
assumes structural parameters similar to that of Draco, with a King core radius 
of $0.18$ kpc and a King limiting radius of $0.93$ kpc~\cite{Odenkirchen:2001pf}. 
Each curve assumes that the
measurement uncertainty on each star is 2 km s$^{-1}$. In both cases, the stars
have been uniformly distributed at projected positions in the galaxies; this provides
a good representation of the present observational configurations.

In addition to their interesting applications for understanding the 
rotation and proper motion of the dSphs, the calculations presented in this section are crucial
for uncovering properties of underlying dark matter distributions. 
For example a strong gradient may reflect ongoing tidal disruption, which would clearly 
affect dark matter mass modeling, as is discussed in more detail in Section~\ref{sec:mass}. 

\section{Proper Motions and Rotation}
\label{sec:tracers}
This section discusses an application of the maximum likelihood 
formalism introduced in Section~\ref{sec:data}, with a specific focus on  
the methodology for extracting an intrinsic rotational signal and proper motions
using an example data set. Extracting rotation from a data set is important for
reasons discussed above, and, in addition to its phenomenological interests,
extracting the proper motions of MW satellites 
may have important implications for understanding the origin of the accretion history of 
MW~\cite{Lynden-Bell1976,D'Onghia:2008bi,Metz:2009ys}. Specifically determining the 
latter would present a unique observational test of MW halo formation within the CDM paradigm. 

Several dSphs have kinematics data sets large enough that 
statistically significant constraints may be placed on the parameters $v_x$, $v_y$, and $A$. For illustrative
purposes this section considers just one example, the Sculptor dSph.  
Sculptor is located at a distance of 80 kpc and has a measured King limiting radius for its stellar distribution 
of $\sim 1.6$ kpc~\cite{Westfall:2005ji}. Given these
parameters it is one of the more spatially extended dSphs. The mass content
of Sculptor has been estimated in several recent papers
~\cite{Battaglia:2008jz,Lokas:2009cp,Strigari:2008ib}, and it 
has been shown that Sculptor may contain some 
degree of rotational support~\cite{Battaglia:2008jz}. Further, 
the previous determinations of the proper motion of Sculptor from its
line-of-sight velocities may indicate a discrepancy between the 
proper motion as determined from this method and from ground and space
based measurements~\cite{Walker:2008ji}. This latter fact may in itself be indicative of
the presence of an intrinsic rotational component, provided the systematics on 
the ground and space-based determinations of the Sculptor proper motions 
are well-understood~\cite{Piatek:2006ks}. 

To extract the rotation and proper motion signal, a simplified model 
is considered by assuming the likelihood function is characterized by 
the six parameters introduced in Section~\ref{sec:data}. It is assumed
that the intrinsic dispersion $\st$ is uniform throughout the galaxy, and
does not depend on any of the parameters of the mass modeling introduced
in Section~\ref{sec:mass} below. Introducing the set of 
parameters discussed in Section~\ref{sec:mass} does not affect the 
reconstruction of the parameters discussed in this section since the
intrinsic dispersion is uncorrelated with the parameters of the function $u$
~\cite{Kaplinghat:2008sm}. 

\begin{figure}[htbp]
\begin{center}
\begin{tabular}{cc}
\includegraphics[height=7cm]{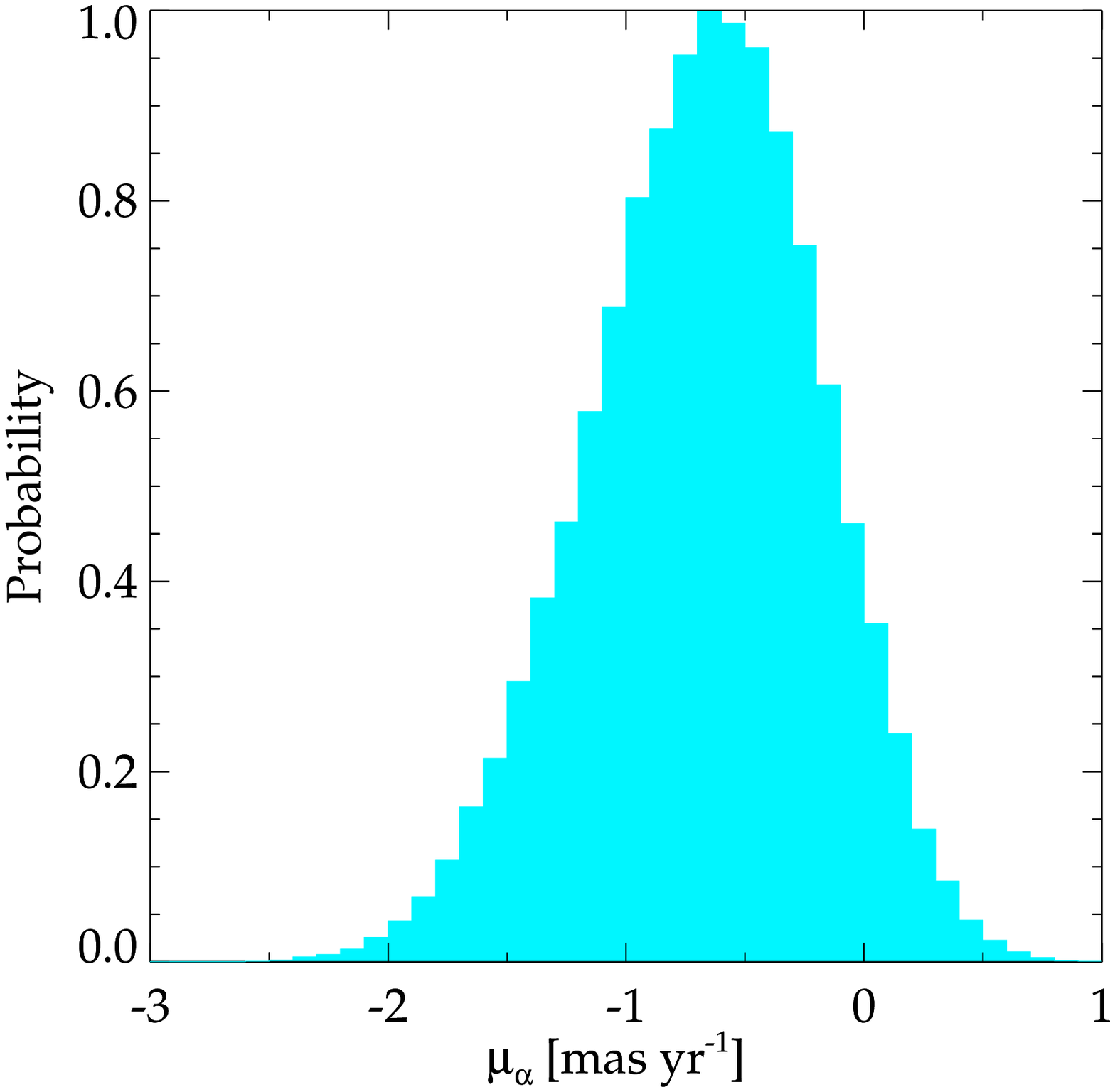} &
\includegraphics[height=7cm]{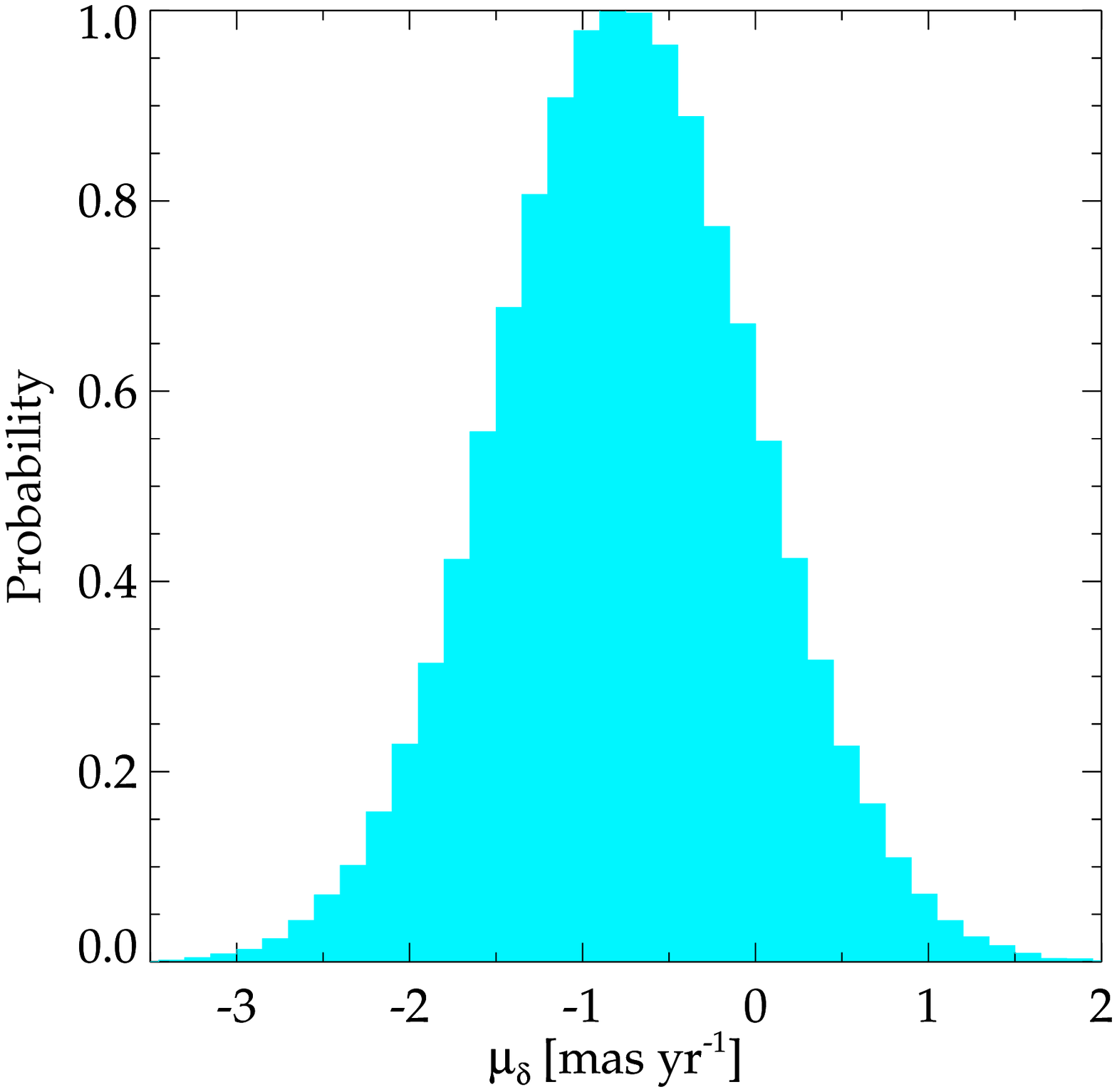} \\
\end{tabular}
\caption{Probability distributions for the proper motions of 
Sculptor using its line-of-sight velocity data. 
The components $\mu_\alpha$ and $\mu_\delta$
give the proper motions in the directions of increasing 
right ascension and declination, respectively.  
\label{fig:propermotions}
}
\end{center}
\end{figure}

In order to determine the probability distributions of the parameters 
$\vec \theta = (v_x,v_y,v_z,A,\phi_0, \sigma_{los})$, a standard
metropolis-hastings algorithm~\cite{Gregory2005}
is used to sample the likelihood function as 
written in Eq.~\ref{eq:likelihood_A}. For all runs described here
$10^4$ accepted points were obtained in each chain, with the first $10\%$ excluded
to account for a conservative burn-in phase. For simplicity, a uniform
proposal distribution is assumed for each of the parameters
over a wide range chosen to encompass physically-accpetable
values for each of these parameters. 
The line-of-sight velocity data used is taken from the Walker 
et al.~\cite{Walker:2008ax} sample, and only those stars with 
$>90\%$ c.l. probably for membership are used in the analysis. 

Figure~\ref{fig:propermotions} shows the posterior probability distributions for
the proper motion components corresponding to $v_x$ and $v_y$, 
$v_x [{\rm km \, s}^{-1}]= 4.74 [D/{\rm pc}] [\mu_\alpha/\mu {\rm as} \, {\rm yr}]$ and 
 $v_y [{\rm km \, s}^{-1}]= 4.74 [D/{\rm pc}] [\mu_\delta/\mu {\rm as}  \, {\rm yr}]$, 
 respectively. The remaining parameters 
$ (v_x,A,\phi_0, \sigma_{los})$ are marginalized over. 
 Though the distributions in Fig.~\ref{fig:propermotions} 
 were obtained by allowing the $A$ parameter to float freely, the distributions
 are found to be relatively unaffected if $A$ is instead fixed so that $A=0$. 
 This reflects the fact that the radial gradient in the velocity of the stars
is distinct from the intrinsic rotational component, which has a sinusoidal 
behavior as a functional of the position angle. The results presented
in Fig.~\ref{fig:propermotions} are in agreement with the measurements of Walker et 
al.~\cite{Walker:2008ji}, though here a larger set of parameters
is marginalized over in this analysis.  

\begin{figure}[htbp]
\begin{center}
\includegraphics[height=8cm]{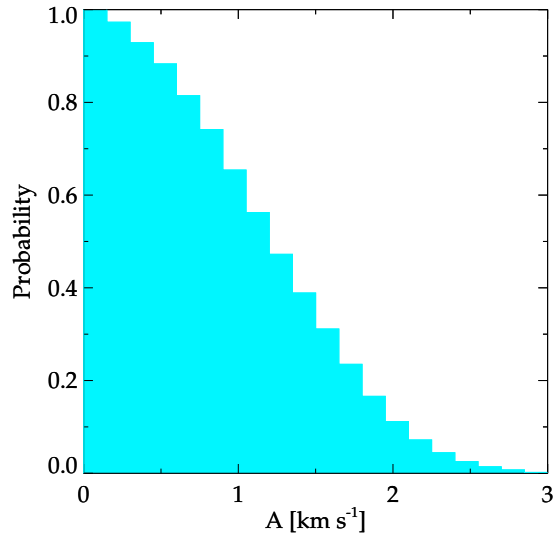}
\caption{The probability for the rotational amplitude, $A$, using the Sculptor 
line-of-sight velocity data. 
\label{fig:A}
}
\end{center}
\end{figure}

Figure~\ref{fig:A} shows the corresponding probability distribution for the
rotational parameter $A$. Again the remaining five parameters 
$ (v_x,v_y,v_z,\phi_0, \sigma_{los})$ are marginalized over. The result
is that, given the rotational parameterization and the using entire distribution of
1352 Sculptor stars, there is no statistically significant detection of rotation. 
From figure~\ref{fig:A} the $90\%$. c.l. upper limit on the rotation 
is found to be $\sim 2$ km s$^{-1}$. The result presented in
figure~\ref{fig:A} is somewhat degenerate with the parameters 
describing $u$; for example if $v_x$ and $v_y$ were (unphysically) set to zero, the
implied upper limit on $A$ reduces by about 50\%. 

Figure~\ref{fig:A} represents the averaged value of $A$ throughout the
entire galaxy. It may be possible that the rotation amplitude in the outer
region differs from the rotation rate in the inner region; if this were the 
case then it is plausible that this effect is washed out in the averaging
process. To provide a simple test for a possible differential rotation rate, 
an additional likelihood analysis was considered with just the outer 
sample of Sculptor stars. Here the outer stars are defined as only
those with projected radius beyond 0.5 kpc. 
Even in this case, there is no 
statistically significant detection of $A$, though in this case
the 90\% c.l. upper limit increases to 10 km s$^{-1}$. 

\section{Mass Distributions and Model Selection Criteria} 
\label{sec:mass}
This section discusses the extension of the maximum likelihood analysis developed in 
Section~\ref{sec:data}, with a goal of using the kinematic data to determine dark matter
mass distributions. 
A calculation of the gravitating mass of a stellar system is one of the more 
fundamental tasks in astronomy, and simple scaling arguments provide some
guidance to anticipate the results. It is worthwhile to first review these arguments
as applied to the dSphs before undertaking a more detailed and model
dependent treatment. 

\subsection{Spherical Mass Modeling}

\subsubsection*{Initial Estimates}-- 
Under the assumption that a star cluster is spherically-symmetric, the orbital 
distribution of the tracer particles are isotropic, that mass follows light, 
and the cluster is isolated 
from any external gravitational potential, the virial theorem provides a mass
estimate of $M_{vir} \simeq r_e \sigma_\star^2/G$, where $r_e$ is the observed
extent of the cluster and $\sigma_\star$ is the velocity dispersion of the stars. 
Although this is probably the simplest estimate one can make for the mass of
a star cluster, it does provide a useful extremum bound. For example Merritt~\cite{Merritt1987}
has show that the virial theorem may be used to derive a lower bound on the mass 
of a star cluster, which is obtained from the assumption that all of the mass is concentrated 
as a point in the center. This minimum mass is given by 
$M_{min} =3 \sigma_\star^2/  \langle r^{-1} \rangle G$, where $1/\langle r^{-1} \rangle$
is the harmonic mean stellar radius in the cluster.  

Of course for dSphs it is not consistent to assume that these systems are isolated, 
since they are orbiting within the extended dark matter halo of the MW. For 
dSphs orbiting with the MW halo, the
minimum mass estimate above is particularly useful, as it in 
turn provides a conservative estimate of the radius at which particles would
be stripped due to the MW potential. As an example consider the case of 
Segue 1, which is a MW satellite with a stellar luminosity $\sim 340$ 
L$_\odot$ at a Galactocentric distance of 28 kpc. From the de-projected
light distribution, the harmonic mean stellar radius is $\sim 10$ pc, and
given the velocity dispersion of $4.3$ km s$^{-1}$~\cite{Geha:2008zr}, 
the implied minimum mass of Segue 1 is $\sim 4 \times 10^5$ M$_\odot$. 
Assuming that Segue 1 is a point mass orbiting in the potential of the 
MW, the radius at which particles would be presently getting stripped is
the Jacobi radius, $r_t = [M /3 M_{\rm MW}]^{1/3} D$, 
where $D = 28$ kpc. Assuming the minimum mass of $M = M_{min}$, 
$r_t \simeq 300$ pc. It is important to note that this provides only 
an estimate of the instantaneous tidal radius; if Segue 1 came 
significantly closer to the MW in the past then this estimate would
differ. The above estimate provides a lower bound on the radius at
which particles would be getting stripped, under the assumption of 
a circular orbit. A similar argument for the tidal radius of Segue 1 
was considered in Geha et al.~\cite{Geha:2008zr} using the 
Illingworth approximation for the mass as $M_{min}$
(For an alternative interpretation for the origin of Segue 1, 
see Ref.~\cite{NiedersteOstholt:2009na}). 

\subsubsection*{Jeans Equation}-- At the next level of detail from the dynamical perspective, an 
estimate for the mass of the dSphs may be obtained 
by appealing to the spherically-symmetric jeans equation, assuming that
the gravitating mass of the system consists of both stars and dark matter. 
The analysis here closely follows the treatment given in the appendix of
Strigari et al.~\cite{Strigari:2008ib}, and refers to this paper for further details.  
A standard discussion of the spherical jeans equations comes
from Ref.~\cite{BT2008}. 

The spherical jeans equation is  
\begin{equation}
\label{eq:jeans}
r \frac{d(\rho_{\star} \sigma_r^2)}{dr} =  - \rho_{\star}(r) V_c^2(r)
        - 2 \beta(r) \rho_{\star} \sigma_r^2.
\end{equation} 
Here $\rho_\star$ is the de-projected stellar density profile, the circular velocity 
is $V_c (r) = GM/r$, and the parameter $\beta (r) = 1 - \sigma_r^2/\sigma_t^2$ 
characterizes the difference between the radial and tangential velocity dispersions
of the stars. 
Integrating $\sigma_r^2$ along the line-of-sight gives the velocity 
dispersion as a function of projected radius, $R$, 
\begin{equation} 
\sigma_{los}^{2}(R) = \frac{2}{I(R)} \int_{R}^{\infty} 
\left ( 1 - \beta \frac{R^{2}}{r^2} \right )
\frac{\rho_{\star} \sigma_{r}^{2} r}{\sqrt{r^2-R^2}} dr. 
\label{eq:sigma}
\end{equation} 
Here, $I(R)$ is the projected surface density of the stellar distribution, and $\rho_\star$ 
is the three-dimensional stellar distribution. In Eq.~(\ref{eq:sigma}), $\sigma_r$ 
depends on the parameterization of the mass distribution of the dark matter component. 
The stellar density profile is taken to be fixed; for example the measurements of
the projected density profiles for many of the classical satellites come from Ref.~\cite{Irwin:1995tb}, 
and more updated profiles from, e.g., Refs.~\cite{Odenkirchen:2001pf,Munoz:2006hx,Smolcic:2007hk,Coleman:2008kk},
while measurements of the density profiles for the ultra-faint satellite come from 
Ref.~\cite{Martin:2008wj}. It is important to note that fixing the stellar density
profile may introduce a degeneracy in determining the projected velocity dispersion profile,
particularly in the central regions~\cite{Zhao:1996mr}. However the effect on the integrated
mass distributions as considered here is less severe, motivating the assumption of fixing $\rho_\star$,
rather than marginalizing over it, in the analysis. 

Given the above assumption for the velocity anisotropy of the stars and for the shape of the dark 
matter profile for the galaxy, the likelihood function can now schematically be written as
\begin{equation} 
{\cal L}[\vec y, \sigma_{los}(\vec \beta, \vec \Phi)] = p[\vec y, \sigma_{los}(\vec \beta, \vec \Phi) | \vec v]. 
\label{eq:likelihood_mass}
\end{equation} 
For compactness, the vector $\vec y = (v_x,v_y,v_z,A,\phi_0)$ has been defined, 
and $\vec \beta$ and $\vec \Phi$ are vectors that describe the stellar velocity anisotropy and
the gravitational potential of the system, respectively. The line-of-sight velocity 
dispersion is dependent on $\vec \beta$ and $\vec \Phi$ through the spherical 
jeans equation. The mass of the system, as well as quantities related to the mass distribution, 
are determined via $\vec \Phi$, and thus by integrating out the model parameters
one may determine the probability distribution for the mass of the system contained within 
a fixed physical radius. 

\subsubsection*{Error Projections on Mass Distribution}--
Before performing an example calculation using Eq.~\ref{eq:likelihood_mass}, 
it is interesting to get an idea as to how the errors on the mass distribution 
depend on the physical radius within which the mass is determined. To perform 
these estimates, we again appeal to the Fisher matrix formalism outlined above. 
However the analysis here is different from above in that now the likelihood depends 
on the vector set of parameters $\vec \beta$ and $\vec \Phi$ in addition to $\vec y$. 

The example considered here uses the velocity data sample from Fornax of 
Walker et al.~\cite{Walker:2008ax}, specifically the stars with $> 90\%$ c.l. for membership. 
This gives a total of 2409 Fornax 
members. The three-dimensional surface density profile for Fornax is assumed
to take the form
\begin{equation} 
\rho_\star(r) \propto \frac{1}{x^a(1+x^b)^{(c-a)/b}} \, e^{-\frac{r^2}{2r_{\rm cut}^2}},
\label{eq:3Dzhao}
\end{equation} 
with the parameters $\{a,b,c,[r_0/{\rm kpc}],[r_{\rm cut}/{\rm kpc}]\} = \{0.3,1.2,3.0,0.8,1.1\}$.
A profile of this form with these parameters is consistent with the recent measurements of Fornax
star counts~\cite{Coleman:2008kk}, though generally 
the results presented are independent of the normalization of the surface density profile. 
The stellar mass-to-light ratio is assumed to be unity, consistent with the 
results presented in Ref.~\cite{Coleman:2008kk}. 
The dark matter density profile is assumed to be the einasto profile, 
\begin{equation}
\ln [\rho(r)/\rho_{-2}] = (-2/\alpha)[(r/r_{-2})^\alpha-1], 
\label{eq:einasto}
\end{equation}
and following CDM simulations, $\alpha=0.17$~\cite{Navarro:2008kc}. The velocity 
anisotropy is assumed to be of the following form, 
\begin{equation}
\beta(r) = (\beta_\infty -\beta_0)r^2/(r_\beta^2+r^2) + \beta_0. 
\label{eq:beta}
\end{equation}
Thus in the Fisher matrix calculation the base set of parameters are now given 
by $\vec \theta = \{\rho_{-2},r_{-2},\beta_0,\beta_1,r_\beta\}$
(the rotational and geometric parameters, $\vec y$ are ignored here: this is justified given that 
the Fornax data is consistent with $A=0$ and that the $\vec y$ parameters
do not correlate with the parameters that determine the mass). 

Given the base set of parameters in $\vec \theta$ used to calculate $\bf F$, 
the error on a derived parameter, $g$, is given by
\begin{equation}
\sigma_g^2 = \sum_{\imath,\jmath} 
\left (\frac{\partial g}{ \partial \theta_\imath} \right)({\bf F^{-1}})_{\imath \jmath}
\left(\frac{\partial g}{ \partial \theta_\jmath} \right). 
\label{fig:fisher_derived}
\end{equation} 
The derived parameter specifically considered here is the log of the mass within a given fixed physical radius
(See Ref.~\cite{Strigari:2007vn} for another example where the derived parameter considered is
the log slope of the dark matter density profile). 
Where desired Gaussian priors may be taken by simply adding $1/\sigma_{aa}^2$ to the $aa$ component of the
Fisher matrix. 
 
\begin{figure}[htbp]
\begin{center}
\includegraphics[height=8cm]{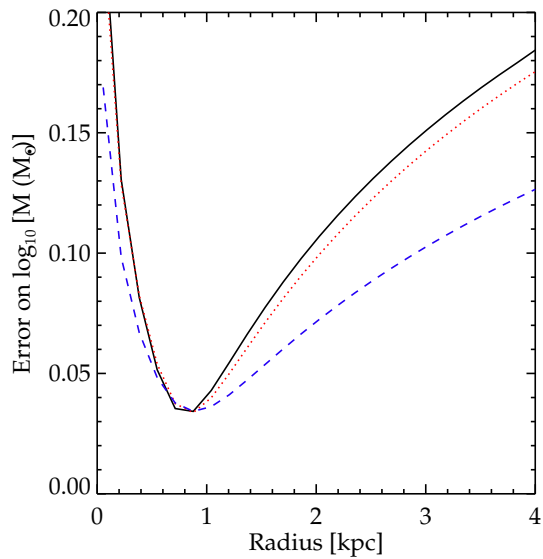}
\caption{Projections for the error on the log of the mass within a given
physical radius for Fornax. The position and the errors 
on the stars from the Walker et al.~\cite{Walker:2008ax} data sample have been used. 
Each curve assumes a different fiducial model for the einasto parameters of the
dark matter halo: $\rho_{-2} = 2 \times 10^7$ M$_\odot$ kpc$^{-3}$ and r$_{-2} = 1$ kpc
(solid black), $\rho_{-2} = 1 \times 10^7$ M$_\odot$ kpc$^{-3}$ and r$_{-2} = 2$ kpc
(dotted red), and $\rho_{-2} = 0.5 \times 10^7$ M$_\odot$ kpc$^{-3}$ and r$_{-2} = 5$ kpc
(dashed blue).  
\label{fig:fisher_mass}
}
\end{center}
\end{figure}

Figure~\ref{fig:fisher_mass} shows the error on the log of mass as a function of 
the physical radius within which the mass is measured. Here  
the fiducial baseline parameters for the velocity anisotropy have been taken as  
$\{\beta_0,\beta_1,r_\beta/{\rm kpc} \} =\{-0.5,0,0.2\}$, implying slightly tangential orbits in the
central region of the halo and isotropic orbits at outer radii. 
Different combinations of $\{\rho_{-2}, r_{-2}\}$
have been taken as indicated to represent the degeneracy between these two parameters when
fitting the data. Each of these parameters sets, combined with an anisotropy model, 
produces a velocity dispersion profile that
roughly fits the profile of Fornax. While the goal here is to not undertake a direct fit to the
data and to explore the exact degeneracy space of these parameters, 
examining these three sets of fiducial parameters gives a feel for
how the constraints on the mass depend on the fiducial parameter set. 
Priors on each of r$_{-2}$ and r$_\beta$ are taken as $1/(5 \, {\rm kpc})^2$, while 
priors on $\beta_0$ and $\beta_1$ are taken as $1/1^2$. Each of these priors are
motivated by the range of these parameters scanned in the algorithm described in the sub-section
below. As is seen, for the stellar profile considered above and
the fiducial set of parameters taken, the best-constrained mass is at a radius $\sim 0.6-1.0$ kpc. 
This best constrained radius is found to be relatively weakly dependent on the 
sets of fiducial parameters, particularly near the best constrained mass, 
provided that they give a good fit to both the star count
and velocity dispersion data of Fornax. For kinematic data sets that have been analyzed, the mass 
is a seen to be strongly-constrained at the approximate half-light radius, which is a 
general property of dispersion supported systems
~\cite{Walker:2009zp,Wolf:2009tu}. 

\subsubsection*{Fornax Mass Distribution}--
The probability for the mass distribution of Fornax is now determined
directly from the kinematic data, and compared 
to the projected error on the mass distribution as determined from
Fig.~\ref{fig:fisher_mass}. 
As above, a metropolis hastings algorithm is used to determine
the respective parameter distributions, and 
the same data for both the star counts and the line-of-sight velocity
distribution have been used. In the parameter  scan, uniform priors have been taken on 
each of the parameters over the following ranges as follows: $\{ \log_{10} [\rho_{-2}/(M_\odot \, {\rm kpc}^{-3})],
r_{-2}/{\rm kpc},\beta_0,\beta_0,r_\beta\}=\{[6:10],[0:10],[-5:1],[-5:1],[0:10]\}$. 

\begin{figure}
\begin{center}
\begin{tabular}{cc}
{\resizebox{6.5cm}{!}{\includegraphics{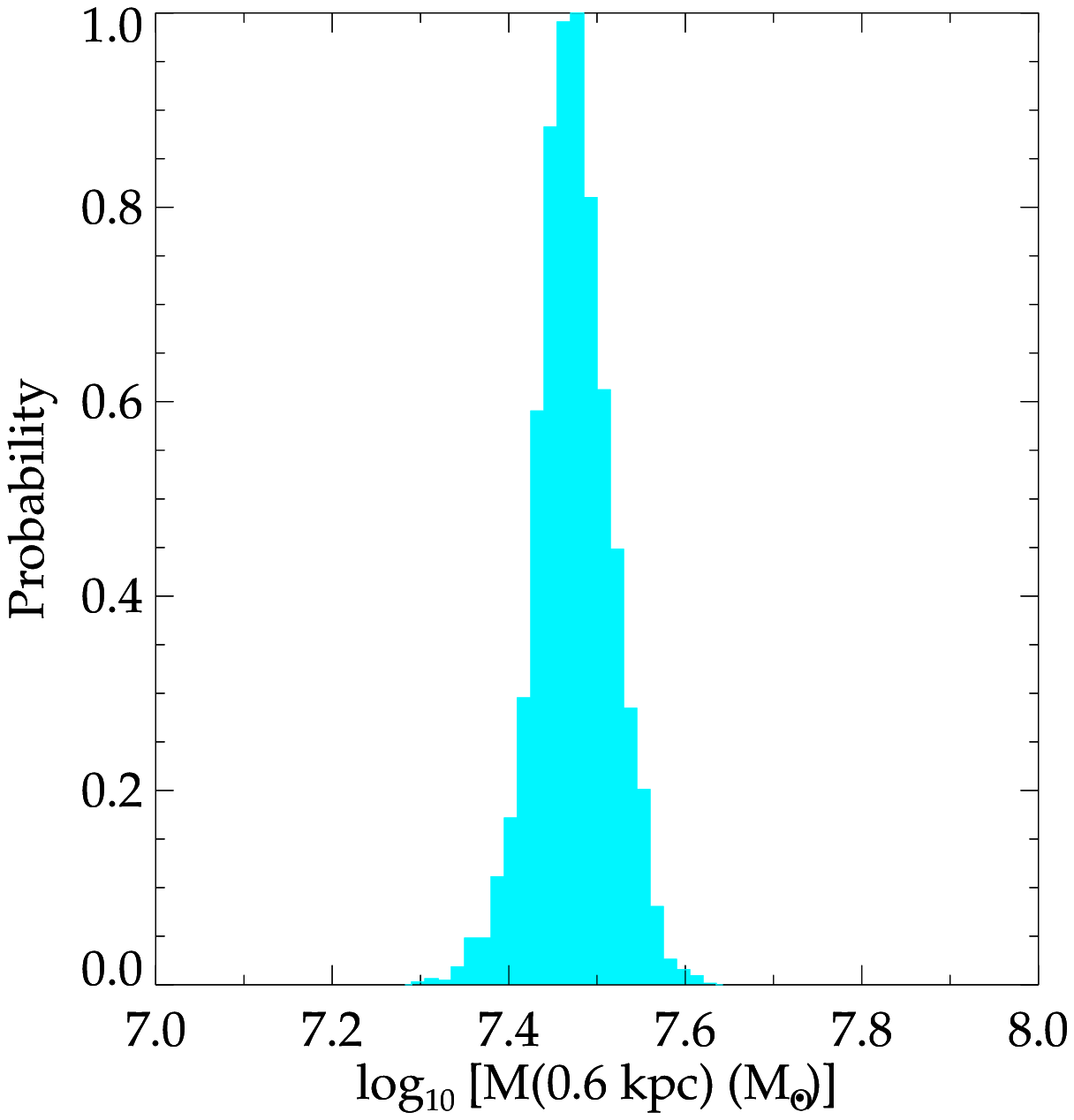}}} &
{\resizebox{6.5cm}{!}{\includegraphics{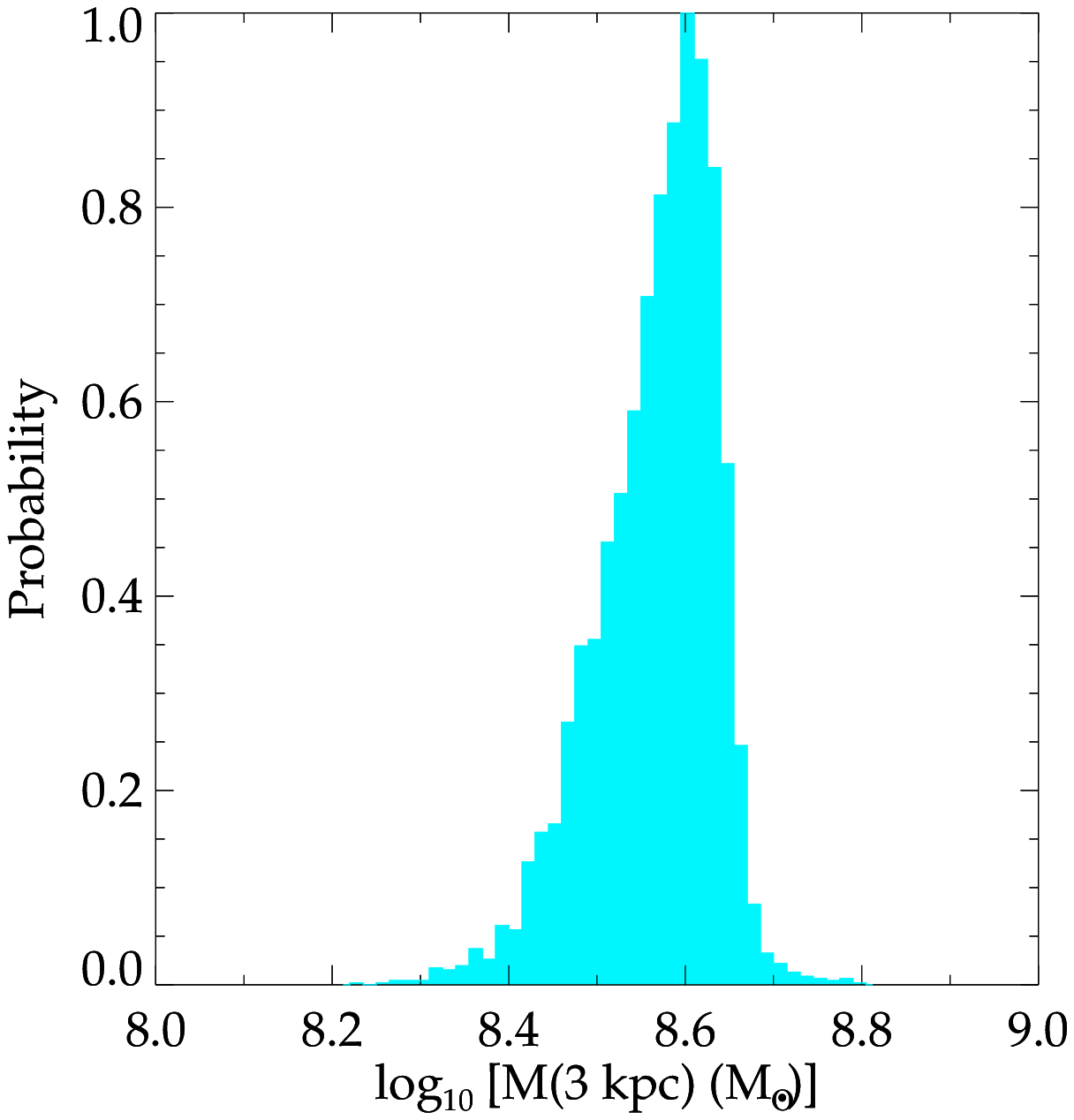}}} \\
\end{tabular}
\end{center}
\caption{The probability distribution for the mass of Fornax within 0.6 kpc ({\em left})
and the mass within its stellar tidal radius ({\em right}), 
defined here to be 3 kpc.  \label{fig:fnx_posteriors}
}
\end{figure}

Figure~\ref{fig:fnx_posteriors} shows two example probability distributions for the Fornax
mass, within 0.6 kpc ({\em left}) and within the approximate Fornax 
stellar tidal radius of 3 kpc ({\em right}). The probability distributions are seen to 
be slightly non-gaussian, particularly the M(3 ${\rm kpc}$) distribution. 
Comparing the approximate width of each of these distributions with the 
errors projected in Fig.~\ref{fig:fisher_mass} provides generally good agreement, 
in spite of the intrinsic assumption in the Fisher matrix formalism that the errors on the parameters
are Gaussian. Specifically for the left panel, a Gaussian fit gives
$\log_{10} [{\rm M}$ (0.6 {\rm kpc})/M$_\odot ] =  7.47 \pm 0.04$. 
These results confirm the general trend seen in Fig.~\ref{fig:fisher_mass}
that the error on the integrated mass within a fixed physical radius increases 
at larger radii towards outer regions of the halo. 

Results of the calculations for the mass distributions of the entire population of dSphs are presented in 
Refs.~\cite{Strigari:2007ma,Strigari:2008ib,Walker:2009zp,Wolf:2009tu}. These results, as well as more recent
determinations, show that the central mass distributions for the
dSphs are very similar, despite an over four order of magnitude variation in their
luminosities. The average density within a spherical radius 
of $\sim 0.3$ kpc is $\sim 0.1$ M$_\odot$ pc$^{-3}$; for the brightest satellites 
baryons can contribute to the potential in this central region, while for the 
least luminous satellites the potential is dominated by dark matter within this
region. Within the context of spherical 
models, these constant central density results are robust to the specific parameterization of the mass distribution, 
primarily due to the fact that the integrated mass is directly constrained via the jeans equation
and the approximately similar scale for the velocity dispersion profiles~\cite{Strigari:2008ib}.  

\subsubsection*{Model Selection}--
The likelihood formalism introduced above does  
not give any information regarding the optimal parameterization of the dark matter 
mass profile. For example referring to the calculation above, is the 
einasto profile with just two free parameters an acceptable description of the data? 
Given the parameterization of the dynamics via the spherical jeans equation, 
we can answer this question and determine how many parameters are required to describe the 
mass profile given the maximum likelihood formalism. Moreover, we can determine
how the parameterization of the density profile depends on the given data sets. 
For example Segue 1, with only 24 measured
line-of-sight velocities, may require a smaller set of parameters than does Fornax, 
which has $\sim 2400$ measured line-of-sight velocities. 

To specifically answer the question 
of how to determine the appropriate set of parameters in maximum likelihood theory
one may appeal to the bayes evidence. For the purposes here the evidence, $E$,  is 
defined as the integral of the likelihood in Eq.~\ref{eq:likelihood_mass} over 
all of the model parameters. When comparing models, the ratio of their 
respective evidences gives an idea of how much more probable one model is
over another. For example if $1 < \Delta \ln E < 2.5$, the different between 
the two models is substantial; for $2.5 < \Delta \ln E < 5$, the different between 
the two models is strong, and for $\Delta \ln E > 5$ the difference between the 
two models is decisive~\cite{Jeffreys}. 

As an illustration, four different dSphs that span a
wide range in their respective number of velocities are considered: 
Segue 1, Sextans, Sculptor, and Fornax. These dSphs have 24, 424, 
1352, and 2409 stars respectively; for the latter three galaxies we consider
only those stars that have a probability of $> 90\%$ membership from
the Walker et al.~\cite{Walker:2008ax} sample. 
For each dSph we determine how many model parameters 
are necessary to describe the data, and we consider several different 
models. 

For the ``Baseline" 3 parameter model, the following range of parameter space is integrated
over for Fornax, Sculptor, and Sextans: 
$\log_{10} [\rho_0/ {\rm M_\odot kpc^{-3}}] = [6:10]$, $\log_{10} [r_0/{\rm kpc}] = [-1:1]$. 
The velocity anisotropy is assumed to be a constant, $\beta$, with a range given by
$\beta\equiv\beta_0=\beta_1=[-2:0.5]$.
For Segue 1 the ranges are the same except for the scale radius, 
which is taken to vary over the range $\log_{10} [r_0/{\rm kpc}] = [-2:0]$; this range is
motivated by the likely upper limit to the dark matter tidal radius for Segue 1
~\cite{Geha:2008zr}. All of the ranges above are chosen as
plausible values to describe the halos of dSphs. A flat prior is chosen over these 
regions; as a further detail one may chose a prior that weights each of these
parameters differently, for by example considering the scatter in the $\rho_s-r_s$
relation as seen in CDM simulations~\cite{Springel:2008cc}.  The bayes evidence 
for the Baseline model will be denoted as $E_0$. 

Three different models are compared to this Baseline 3 parameter model: i) a model in which 
the parameter space for $\alpha$ is enlarged in the range $[0.14:0.3]$
(corresponding to $1/\alpha \simeq [3-7]$), 
so that central slopes that are both more flat and more
steep than the CDM value are allowed; 
ii) a model with the Baseline 3-parameter $\rho(r)$ profile, but with a three-parameter velocity 
anisotropy profile which depends on the three-dimensional physical radius as in
Eq.~\ref{eq:beta}
and iii) a model in which $\alpha = [0.14:0.3]$ and the $\beta(r)$ profile in 
Eq.~\ref{eq:beta} is assumed. 

Model i) is thus described by four parameters, while model ii) is described by 
five free parameters, and model iii) is described by six free parameters. 
In Table~\ref{tab:table1}, we define model i) as the ``Exp" 
model, model ii) is denoted as the ``$\vec \beta$" model, and model iii)
is denoted as ``$\vec \beta +$ EXP". 
These models provide useful illustrations of the calculation of the evidence as applied to
dSphs; alternative models may of course be defined and even larger parameter spaces
may be explored. The utility of the above models as defined allows us to explore to what
extent CDM-like inner slopes are more favored, and to what extent 
an alternative parameterization of the velocity anisotropy provides a better fit to the data
as compared to simply changing the value of the inner slope. 

The results for the ratio of the bayes evidence for the various models, relative to $E_0$,
are shown in Table~\ref{tab:table1}. For each of the galaxies, we see roughly the same
pattern; as more parameters are added, the better that the model fits the data. 
This result implies that models with larger sets of 
parameters are favored even after penalization for the larger volume of parameter
space that is integrated over. Allowing for a larger volume of parameter space  
for the dark matter density profile affects the evidence more than simply varying 
the shape of the anisotropy profile. In total, the best fitting models are those that
allow both the velocity anisotropy and the central slope to vary freely, i.e. the $\vec \beta +$ EXP
models. 

The results in Table~\ref{tab:table1} indicate that for all four galaxies 
variable velocity anisotropies are slightly preferred relative to those with constant velocity anisotropy, 
and that central dark matter profiles both less cuspy and more cuspy than
$\Lambda$CDM based fits are equally acceptable. Future data sets, both line-of-sight velocities
and potential proper motion measurements for stars in dSphs
~\cite{Wilkinson:2001ut,Strigari:2007vn,Majewski:2009zf}, will be important in narrowing the acceptable 
ranges for both the velocity anisotropy and the central slope. 

\begin{table*}
\centering
\begin{tabular}{|c||c|c|c|c|}
\hline
dSph & Segue 1& Sextans&Sculptor & Fornax \\   
\hline 
\hline 
\# of stars & 24 & 424&1352&2409\\
\hline
Exp &3.4&3.3&3.3&3.6\\ 
\hline 
$\vec \beta$&1.6&0.9&0.8&1.3\\ 
\hline
$\vec \beta$ + Exp &5.0&4.3&4.1&4.9\\ 
\hline 
\end{tabular}
\caption{\label{tab:table1} Parameter ranges and evidences for various models.
The columns are for different dSphs, and the rows are for different models as
described in the text. Each entry in the table gives the ratio of the Evidence 
for the given model, $E_{\rm model}$, 
with respect to the Baseline 3 parameter model, $E_0$, defined in the text, 
$\ln(E_{\rm model}/E_0)$. 
}
\end{table*}

\section{Conclusion}
\label{sec:conclusion}
This article has discussed the analysis of kinematic data from Milky Way 
dwarf spheroidals, with a primary motivation of 1) understanding physical
quantities that are well-constrained by the data and 2) understanding the systematics 
that underly the determination of the dark matter masses of these systems, given the 
simplest assumption that the dSphs are purely pressure supported systems. 
Of the possible systematics perhaps the most significant and observationally-
accessible is the determination of a velocity gradient in
the data sample, which may be indicative of 
tidal disruption from the potential well of the Milky Way. 
The results in 
the literature indicate that, based on the kinematic data alone, 
velocity gradients due to tidal disruption or rotation 
are not conclusively present in any of the dSphs. This article has 
provided an example, using a simple parameterization, of how to 
search for rotation in the kinematic data sets using a maximum likelihood
analysis. The kinematic sample of Sculptor was analyzed, and it was
found that the maximum likelihood rotational amplitude is zero, with
an upper limit of $\sim 2$ km s$^{-1}$ at 90\% c.l. The magnitude of
these errors are consistent with the projected magnitude of the errors
from theoretical modeling. 

When modeling the mass distribution of the dark matter halos of the dSphs, 
degeneracies between model parameters affect the determination of the
total mass profiles, even in the context of the simplest spherical models. To 
shed light on these degeneracies, this article has discussed a new 
criteria for model selection applied to the dSph kinematic data sets, 
taking a step towards determining how many parameters
are needed to describe the mass distribution of spherical halos. For the four dSphs
studied here, chosen because they have a wide range of available line-of-sight
velocities, it is shown that, assuming CDM-motivated Einasto profiles for the
dark matter halos, models with variable velocity anisotropy are slightly 
preferred relative to those with constant velocity anisotropy. Further, central
slopes for the dark matter profile that are found in CDM simulations are 
not a unique description of the data sets; both more cuspy and
less cuspy models are allowed for the central slope. This is primarily due
to the degeneracy between the central dark matter slope with the 
central stellar profile and the velocity anisotropy distribution
~\cite{Strigari:2007vn}. 

Future photometric and kinematic data sets promise to further pin down 
the mass distributions of the dSph dark matter halos. Upcoming data for
the ultra-faint satellites will be particularly important, and may be able to 
show whether any tidal effects are present in these galaxies. Further, 
development of non-spherical distributions for both the light and dark matter
should be considered given these data sets (for initial results along these
lines see Ref~\cite{Lokas:2009ty}). 
Controlling systematics in these data sets will prove to be important step 
towards further testing the currently favored $\Lambda$CDM theory
of structure formation. 

\section*{Acknowledgments} 
I am grateful to Andrey Kravtsov, Matt Walker, and Beth Willman for discussion and comments on this article. 
I additionally acknowledge the anonymous referee for constructive comments
and critique that improved the content and presentation. 
Support for this work was provided by NASA through Hubble Fellowship grant 
HF-01225.01 awarded by the Space Telescope Science Institute, which is 
operated by the Association of Universities for Research in Astronomy, Inc., 
for NASA, under contract NAS 5-26555.

\vspace{2cm}

\bibliography{paper}

\begin{thebibliography}{10}

\bibitem{Shapley}
H.~{Shapley}.
\newblock {A Stellar System of a New Type}.
\newblock {\em Harvard College Observatory Bulletin}, 908:1--11, March 1938.

\bibitem{Mateo:1998wg}
M.~Mateo.
\newblock {Dwarf Galaxies of the Local Group}.
\newblock {\em Ann. Rev. Astron. Astrophys.}, 36:435--506, 1998.

\bibitem{Aaronson1983}
M.~{Aaronson}.
\newblock {Accurate radial velocities for carbon stars in Draco and Ursa Minor
  - The first hint of a dwarf spheroidal mass-to-light ratio}.
\newblock {\em Astrophys. J.}, 266:L11--L15, March 1983.

\bibitem{Suntzeff1993}
N.~B. {Suntzeff}, M.~{Mateo}, D.~M. {Terndrup}, E.~W. {Olszewski},
  D.~{Geisler}, and W.~{Weller}.
\newblock {Spectroscopy of Giants in the Sextans Dwarf Spheroidal Galaxy}.
\newblock {\em Astrophys. J.}, 418:208--+, November 1993.

\bibitem{Mateo1991}
M.~{Mateo}, E.~{Olszewski}, D.~L. {Welch}, P.~{Fischer}, and W.~{Kunkel}.
\newblock {A kinematic study of the Fornax dwarf spheroidal galaxy}.
\newblock {\em Astrophys. J.}, 102:914--926, September 1991.

\bibitem{Mateo1993}
M.~{Mateo}, E.~W. {Olszewski}, C.~{Pryor}, D.~L. {Welch}, and P.~{Fischer}.
\newblock {The Carina dwarf spheroidal galaxy - How dark is it?}
\newblock {\em Astrophys. J}, 105:510--526, February 1993.

\bibitem{Lake1990}
G.~{Lake}.
\newblock {The distribution of dark matter in Draco and Ursa Minor}.
\newblock {\em Mon. Not. Roy. Astron. Soc.}, 244:701--705, June 1990.

\bibitem{Gerhard1992}
O.~E. {Gerhard} and D.~N. {Spergel}.
\newblock {Dwarf spheroidal galaxies and the mass of the neutrino}.
\newblock {\em Astrophys. J.}, 389:L9--L11, April 1992.

\bibitem{Faber1983}
S.~M. {Faber} and D.~N.~C. {Lin}.
\newblock {Is there nonluminous matter in dwarf spheroidal galaxies}.
\newblock {\em Astrophys. J.}, 266:L17--L20, March 1983.

\bibitem{Gilmore2007}
G.~{Gilmore}, M.~I. {Wilkinson}, R.~F.~G. {Wyse}, J.~T. {Kleyna}, A.~{Koch},
  N.~W. {Evans}, and E.~K. {Grebel}.
\newblock {The Observed Properties of Dark Matter on Small Spatial Scales}.
\newblock {\em Astrophys. J.}, 663:948--959, July 2007.

\bibitem{Walker2007}
M.~G. {Walker}, M.~{Mateo}, E.~W. {Olszewski}, O.~Y. {Gnedin}, X.~{Wang},
  B.~{Sen}, and M.~{Woodroofe}.
\newblock {Velocity Dispersion Profiles of Seven Dwarf Spheroidal Galaxies}.
\newblock {\em Astrophys. J}, 667:L53--L56, September 2007.

\bibitem{Walker:2008ji}
M.~G. Walker, M.~Mateo, and E.~W. Olszewski.
\newblock {Systemic Proper Motions of Milky Way Satellites from Stellar
  Redshifts: The Carina, Fornax, Sculptor, and Sextans Dwarf Spheroidals}.
\newblock {\em Astrophys. J.}, 688:L75--L78, December 2008.

\bibitem{Strigari:2008ib}
L.~E. Strigari et~al.
\newblock {A common mass scale for satellite galaxies of the Milky Way}.
\newblock {\em Nature}, 454:1096--1097, 2008.

\bibitem{Lokas:2009cp}
E.~L. Lokas.
\newblock {The mass and velocity anisotropy of the Carina, Fornax, Sculptor and
  Sextans dwarf spheroidal galaxies}.
\newblock {\em Mon. Not. Roy. Astron. Soc.}, 394:L102--L106, March 2009.

\bibitem{Willman:2004kk}
B.~Willman et~al.
\newblock {A New Milky Way Companion: Unusual Globular Cluster or Extreme Dwarf
  Satellite?}
\newblock {\em Astron. J}, 129:2692--2700, June 2005.

\bibitem{Willman:2005cd}
B.~Willman et~al.
\newblock {A New Milky Way Dwarf Galaxy in Ursa Major}.
\newblock {\em Astrophys. J.}, 626:L85--L88, 2005.

\bibitem{Belokurov:2006ph}
V.~Belokurov et~al.
\newblock {Cats and Dogs, Hair and A Hero: A Quintet of New Milky Way
  Companions}.
\newblock {\em Astrophys. J.}, 654:897--906, 2007.

\bibitem{Munoz:2006vg}
R.~R. Munoz et~al.
\newblock {Exploring Halo Substructure with Giant Stars: The Dynamics and
  Metallicity of the Dwarf Spheroidal in Bootes}.
\newblock {\em Astrophys. J.}, 650:L51--L54, 2006.

\bibitem{Martin:2007ic}
N.~F. Martin, R.~A. Ibata, S.~C. Chapman, M.~Irwin, and G.~F. Lewis.
\newblock {A Keck/DEIMOS spectroscopic survey of faint Galactic satellites:
  searching for the least massive dwarf galaxies}.
\newblock {\em Mon. Not. Roy. Astron. Soc.}, 380:281--300, September 2007.

\bibitem{Simon:2007dq}
J.~D. Simon and M.~Geha.
\newblock {The Kinematics of the Ultra-Faint Milky Way Satellites: Solving the
  Missing Satellite Problem}.
\newblock {\em Astrophys. J.}, 670:313--331, 2007.

\bibitem{Geha:2008zr}
M.~Geha et~al.
\newblock {The Least Luminous Galaxy: Spectroscopy of the Milky Way Satellite
  Segue 1}.
\newblock {\em Astrophys. J.}, 692:1464--1475, 2009.

\bibitem{Kirby2008}
E.~N. {Kirby}, J.~D. {Simon}, M.~{Geha}, P.~{Guhathakurta}, and A.~{Frebel}.
\newblock {Uncovering Extremely Metal-Poor Stars in the Milky Way's Ultrafaint
  Dwarf Spheroidal Satellite Galaxies}.
\newblock {\em Astrophys. J.}, 685:L43--L46, September 2008.

\bibitem{Strigari:2007ma}
L.~E. Strigari et~al.
\newblock {Redefining the Missing Satellites Problem}.
\newblock {\em Astrophys. J.}, 669:676--683, November 2007.

\bibitem{Strigari:2006rd}
L.~E. Strigari, S.~M. Koushiappas, J.~S. Bullock, and M.~Kaplinghat.
\newblock {Precise constraints on the dark matter content of Milky Way dwarf
  galaxies for gamma-ray experiments}.
\newblock {\em Phys. Rev.}, D75:083526, 2007.

\bibitem{Essig:2009jx}
R.~Essig, N.~Sehgal, and L.~E. Strigari.
\newblock {Bounds on Cross-sections and Lifetimes for Dark Matter Annihilation
  and Decay into Charged Leptons from Gamma-ray Observations of Dwarf
  Galaxies}.
\newblock {\em Phys. Rev.}, D80:023506, 2009.

\bibitem{Martinez:2009jh}
G.~D. Martinez, J.~S. Bullock, M.~Kaplinghat, L.~E. Strigari, and R.~Trotta.
\newblock {Indirect Dark Matter Detection from Dwarf Satellites: Joint
  Expectations from Astrophysics and Supersymmetry}.
\newblock {\em JCAP}, 0906:014, 2009.

\bibitem{Hogan:2000bv}
C.~J. Hogan and J.~J. Dalcanton.
\newblock {New dark matter physics: Clues from halo structure}.
\newblock {\em Phys. Rev.}, D62:063511, 2000.

\bibitem{Gerhard1991}
O.~E. {Gerhard}.
\newblock {A new family of distribution functions for spherical galaxies}.
\newblock {\em Mon. Not. Roy. Astron. Soc.}, 250:812--830, June 1991.

\bibitem{Gregory2005}
P.~C. {Gregory}.
\newblock {\em {Bayesian Logical Data Analysis for the Physical Sciences: A
  Comparative Approach with `Mathematica' Support}}.
\newblock Cambridge University Press, 2005.

\bibitem{Strigari:2007at}
L.~E. {Strigari}, S.~M. {Koushiappas}, J.~S. {Bullock}, M.~{Kaplinghat}, J.~D.
  {Simon}, M.~{Geha}, and B.~{Willman}.
\newblock {The Most Dark-Matter-dominated Galaxies: Predicted Gamma-Ray Signals
  from the Faintest Milky Way Dwarfs}.
\newblock {\em Astrophys. J}, 678:614--620, May 2008.

\bibitem{KS:unpub}
M.~Kaplinghat and L.E. Strigari.
\newblock unpublished.

\bibitem{Kaplinghat:2008sm}
M.~Kaplinghat and L.~E. Strigari.
\newblock {Proper Motion of Milky Way Dwarf Spheroidals from Line-of-Sight
  Velocities}.
\newblock {\em Astrophys. J}, 682:L93--L96, August 2008.

\bibitem{vanderMarel:2002kq}
R.~P. van~der Marel et~al.
\newblock {New Understanding of Large Magellanic Cloud Structure, Dynamics and
  Orbit from Carbon Star Kinematics}.
\newblock {\em Astron. J.}, 124:2639--2663, 2002.

\bibitem{vanderMarel:2007yw}
R.~P. {van der Marel} and P.~{Guhathakurta}.
\newblock {M31 Transverse Velocity and Local Group Mass from Satellite
  Kinematics}.
\newblock {\em Astrophys. J}, 678:187--199, May 2008.

\bibitem{Martin:2008wj}
N.~F. {Martin}, J.~T.~A. {de Jong}, and H.-W. {Rix}.
\newblock {A Comprehensive Maximum Likelihood Analysis of the Structural
  Properties of Faint Milky Way Satellites}.
\newblock {\em Astrophys. J.}, 684:1075--1092, September 2008.

\bibitem{Odenkirchen:2001pf}
M.~Odenkirchen et~al.
\newblock {New insights on the Draco dwarf spheroidal galaxy from SDSS: a
  larger radius and no tidal tails}.
\newblock {\em Astron. J.}, 122:2538--2553, 2001.

\bibitem{Lynden-Bell1976}
D.~{Lynden-Bell}.
\newblock {Dwarf galaxies and globular clusters in high velocity hydrogen
  streams}.
\newblock {\em Mon. Not. Roy. Astron. Soc.}, 174:695--710, March 1976.

\bibitem{D'Onghia:2008bi}
E.~D'Onghia and G.~Lake.
\newblock {Small Dwarf Galaxies within Larger Dwarfs: Why Some Are Luminous
  while Most Go Dark}.
\newblock {\em Astrophys. J.}, 686:L61--L65, October 2008.

\bibitem{Metz:2009ys}
M.~Metz, P.~Kroupa, C.~Theis, G.~Hensler, and H.~Jerjen.
\newblock {Did the Milky Way Dwarf Satellites Enter The Halo as a Group?}
\newblock {\em Astrophys. J}, 697:269--274, May 2009.

\bibitem{Westfall:2005ji}
K.~B. Westfall et~al.
\newblock {Exploring Halo Substructure with Giant Stars VIII: The Extended
  Structure of the Sculptor Dwarf Spheroidal Galaxy}.
\newblock {\em Astron. J.}, 131:375--406, 2006.

\bibitem{Battaglia:2008jz}
G.~Battaglia et~al.
\newblock {The Kinematic Status and Mass Content of the Sculptor Dwarf
  Spheroidal Galaxy}.
\newblock {\em Astrophys. J}, 681:L13--L16, July 2008.

\bibitem{Piatek:2006ks}
S.~Piatek et~al.
\newblock {Proper Motions of Dwarf Spheroidal Galaxies from Hubble Space
  Telescope Imaging. IV: Measurement for Sculptor}.
\newblock {\em Astron. J.}, 133:818--844, 2007.

\bibitem{Walker:2008ax}
M.~G. Walker, M.~Mateo, and E.~Olszewski.
\newblock {Stellar Velocities in the Carina, Fornax, Sculptor and Sextans dSph
  Galaxies: Data from the Magellan/MMFS Survey}.
\newblock {\em Astron. J}, 137:3100--3108, February 2009.

\bibitem{Merritt1987}
D.~{Merritt}.
\newblock {The distribution of dark matter in the coma cluster}.
\newblock {\em Astrophys. J}, 313:121--135, February 1987.

\bibitem{NiedersteOstholt:2009na}
M.~{Niederste-Ostholt}, V.~{Belokurov}, N.~W. {Evans}, G.~{Gilmore}, R.~F.~G.
  {Wyse}, and J.~E. {Norris}.
\newblock {The origin of Segue 1}.
\newblock {\em Mon. Not. Roy. Astron. Soc.}, 398:1771--1781, October 2009.

\bibitem{BT2008}
J.~{Binney} and S.~{Tremaine}.
\newblock {\em {Galactic Dynamics: Second Edition}}.
\newblock Princeton University Press, 2008.

\bibitem{Irwin:1995tb}
M.~Irwin and D.~Hatzidimitriou.
\newblock {Structural parameters for the Galactic dwarf spheroidals}.
\newblock {\em Mon. Not. Roy. Astron. Soc.}, 277:1354--1378, 1995.

\bibitem{Munoz:2006hx}
R.~R. Munoz et~al.
\newblock {Exploring Halo Substructure with Giant Stars XI: The Tidal Tails of
  the Carina Dwarf Spheroidal and the Discovery of Magellanic Cloud Stars in
  the Carina Foreground}.
\newblock {\em Astrophys. J.}, 649:201--223, 2006.

\bibitem{Smolcic:2007hk}
V.~Smolcic et~al.
\newblock {Improved photometry of SDSS crowded field images: Structure and dark
  matter content in the dwarf spheroidal galaxy Leo I}.
\newblock {\em Astron. J.}, 134:1901--1915, 2007.

\bibitem{Coleman:2008kk}
M.~G. {Coleman} and J.~T.~A. {de Jong}.
\newblock {A Deep Survey of the Fornax dSph. I. Star Formation History}.
\newblock {\em Astrophys. J}, 685:933--946, October 2008.

\bibitem{Zhao:1996mr}
H.-S. Zhao.
\newblock {Analytical dynamical models for double power-law galactic nuclei}.
\newblock {\em Mon. Not. Roy. Astron. Soc.}, 287:525--537, May 1997.

\bibitem{Navarro:2008kc}
J.~F. Navarro et~al.
\newblock {The Diversity and Similarity of Cold Dark Matter Halos}.
\newblock arXiv:0810.1522, 2008.

\bibitem{Strigari:2007vn}
L.~E. Strigari, J.~S. Bullock, and M.~Kaplinghat.
\newblock {Determining the Nature of Dark Matter with Astrometry}.
\newblock {\em Astrophys. J.}, 657:L1--L4, 2007.

\bibitem{Walker:2009zp}
M.~G. {Walker}, M.~{Mateo}, E.~W. {Olszewski}, J.~{Pe{\~n}arrubia}, N.~{Wyn
  Evans}, and G.~{Gilmore}.
\newblock {A Universal Mass Profile for Dwarf Spheroidal Galaxies?}
\newblock {\em Astrophys. J.}, 704:1274--1287, October 2009.

\bibitem{Wolf:2009tu}
J.~Wolf et~al.
\newblock {Accurate Masses for Dispersion-supported Galaxies}.
\newblock arXiv:0908.2995, 2009.

\bibitem{Jeffreys}
H.~{Jeffreys}.
\newblock {\em {Theory of Probability}}.
\newblock Oxford University Press, 1961.

\bibitem{Springel:2008cc}
V.~{Springel} et~al.
\newblock {The Aquarius Project: the subhaloes of galactic haloes}.
\newblock {\em Mon. Not. Roy. Astron. Soc.}, 391:1685--1711, December 2008.

\bibitem{Wilkinson:2001ut}
M.~I. Wilkinson, J.~Kleyna, N.~W. Evans, and G.~Gilmore.
\newblock {Dark Matter in Dwarf Spheroidals I: Models}.
\newblock {\em Mon. Not. Roy. Astron. Soc.}, 330:778, 2002.

\bibitem{Majewski:2009zf}
S.~R. Majewski et~al.
\newblock {Galactic Dynamics and Local Dark Matter}.
\newblock arXiv:0902.2759, 2009.

\bibitem{Lokas:2009ty}
E.~L. Lokas, S.~Kazantzidis, J.~Klimentowski, L.~Mayer, and S.~Callegari.
\newblock {The stellar structure and kinematics of dwarf spheroidal galaxies
  formed by tidal stirring}.
\newblock arXiv:0906.5084, 2009.

\end{thebibliography}

\end{document}